\documentclass[aps,twocolumn,showpacs]{revtex4}
\usepackage{footnote}
\usepackage{amsmath}
\usepackage{amssymb}
\usepackage{verbatim}
\usepackage{graphicx}
\usepackage{array}

\begin{document}

\title{Impact of Colored Environmental Noise on the Extinction of a Long-Lived Stochastic Population: Role of the Allee Effect}
\author{Eitan Y. Levine}
\email{eitan.abc@gmail.com}
\author{Baruch Meerson}
\email{meerson@cc.huji.ac.il}
\affiliation{Racah Institute of Physics, Hebrew University of Jerusalem, Jerusalem 91904, Israel}

\begin{abstract}
We study the combined impact of a colored environmental noise and demographic noise on the extinction risk of a long-lived and well-mixed isolated stochastic population which exhibits the Allee effect.
The environmental noise modulates the population birth and death rates.
Assuming that the Allee effect is strong, and the environmental noise is positively correlated and Gaussian, we derive a Fokker-Planck equation for the joint probability distribution of the population sizes and environmental fluctuations.
In WKB approximation this equation reduces to an effective two-dimensional Hamiltonian mechanics, where the most likely path to extinction and the most likely environmental fluctuation are encoded in an instanton-like trajectory in the phase space.
The mean time to extinction $\tau$ is related to the mechanical action along this trajectory.
We obtain new analytic results for short-correlated, long-correlated and relatively weak environmental noise.
The population-size dependence of $\tau$ changes from exponential for weak environmental noise to no dependence for strong noise, implying a greatly increased extinction risk.
The theory is readily extendable to population switches between different metastable states, and to stochastic population explosion, due to a combined action of demographic and environmental noise.
\end{abstract}


\maketitle

\newcommand{\pdf}{\mathcal{P}}                         
\newcommand{\pder}[2]{\frac{\partial #1}{\partial #2}} 
\newcommand{\eps}{\varepsilon}                         
\newcommand{\ie}{\textit{i.e.}}                        
\newcommand{\eg}{\textit{e.g.}}                        
\newcommand{\lt}{\left}                                
\newcommand{\rt}{\right}                               
\newcommand{\ra}{\rightarrow}                          
\newcommand{\xra}{\xrightarrow}                        


\section{Introduction}
A long-lived isolated stochastic population ultimately goes extinct via a large fluctuation: an unusual chain of deleterious events resulting from the demographic noise (the intrinsic discreteness of individuals and random nature of birth-death processes) and environmental variations, see Ref.~\cite{OM} for a recent review.
It is important to understand how the interplay of environmental and demographic noises determines the mean time to extinction (MTE) \cite{color_rev}.
Early models postulated that the environmental noise, which modulates the birth and death rates of the population, is delta-correlated in time \cite{Leigh,Lande}.
Later on, population biologists realized, mostly via stochastic simulations, that temporal autocorrelation, or color, of environmental noise may have a considerable effect on population extinction \cite{OM,color_rev}.
These insights inspired physicists who developed a theoretical framework for the analysis of a joint action of demographic and colored environmental noise on extinction of an established population  whose dynamics follows a simple stochastic logistic model \cite{KMS}.
This theoretical framework provided a transparent way of evaluating the MTE and finding the optimal environmental fluctuation that determines the optimal (most likely) path of the population to extinction.
The theory of \cite{KMS} predicted the MTE in different regions of a two-dimensional ``phase diagram" whose axes are the properly rescaled intensity (or, alternatively, variance), and the correlation time of the environmental noise.
It tracked how the population-size dependence of the MTE changes from exponential with no environmental noise to a power law for a short-correlated noise and to no dependence for long-correlated noise.
It also established the validity domains of the white-noise limit and adiabatic limit.
(In the adiabatic limit the environmental noise is assumed to vary very slowly compared with the relaxation rate of the population toward the attracting fixed point of the deterministic rate equation.)

The simple logistic models adopted in Refs.~\cite{Leigh,Lande,KMS} do not account for the demographic Allee effect, by which population biologists mean a host of effects leading to an effective reduction in the per-capita growth rate at small population size \cite{Allee}.
When the Allee effect is significant, a non-zero critical population size for establishment arises.
If the initial population size is smaller than the critical size, the population quickly goes extinct.
If the initial population size is greater than the critical one, an established population appears.
Population biologists have argued that the Allee effect may influence, in a significant way, the population extinction risk due to the demographic and environmental noise \cite{Ripa}.
No satisfactory theoretical framework, however, has been developed.

The present work attempts to close this gap.
We formulate a minimal theoretical framework for this problem by considering a simple set of stochastic reactions which mimics the Allee effect in a well-mixed population.
The per-capita rates are modulated by a positively correlated Gaussian noise with given magnitude and correlation time.
We assume that the Allee effect is so strong, that the established population size, as predicted by the deterministic rate equation, is close to the critical population size for establishment.
In this limit (that is, close to the saddle-node bifurcation of the deterministic rate equation) a Fokker-Planck equation can be derived, which accurately describes the time evolution of the joint probability distribution of the population sizes and environmental fluctuations.
Throughout this work we assume that both the environmental noise and the demographic noise are weak, so the MTE of the population is very long compared with the characteristic relaxation time predicted by the (noiseless) deterministic rate equation for this population.
This enables us to use a small-noise approximation due to Freidlin and Wentzell \cite{FW98}: essentially,  a dissipative variant of WKB approximation. 
The WKB approximation reduces the Fokker-Planck equation to an effective two-dimensional classical mechanics.
The optimal path of the population to extinction and the optimal environmental fluctuation are encoded in an instanton-like
trajectory in the Hamiltonian phase space of this classical mechanics, while the MTE is related to the mechanical action along the instanton.

We solve the effective mechanical problem, and obtain analytic estimates for the MTE, perturbatively in three different limits: of short-correlated, long-correlated, and relatively weak environmental noise, for a population exhibiting a strong Allee effect.
We also find, in each of these limits, the optimal (most likely) path of the population to extinction and the optimal environmental fluctuation.
We complement our analytic results by solving numerically the equations of motion of the effective classical mechanics. We find that the Allee effect has a strong impact on the MTE.
It was discovered more than 30 years ago by Leigh \cite{Leigh,Lande} that, without the Allee effect, a strong uncorrelated (white) environmental noise changes the population-size dependence of the MTE from an exponential to a power-law with a large exponent.
We show here that, in the presence of the strong Allee effect, no power law appears.
Here the population-size dependence of the MTE changes from exponential for weak environmental noise to no dependence for strong environmental noise.
Our theory is readily extendable to population  \emph{switches} between different metastable states, and to noise-induced population explosion,  due to a combined action of demographic and environmental noise.
Where possible, we compare our results with previous ones.

We reiterate that both demographic and environmental noises are weak in our theory.
Therefore, when we call the environmental noise weak or strong, we only mean that it is weak or strong compared with the demographic noise.

The outline of the paper is as follows.
Sections II and III include preliminaries.
In Section II we introduce a simple model of long-lived stochastic population which exhibits the Allee effect and ultimately goes extinct because of demographic noise.
We start with the deterministic limit of the model and then outline its stochastic behavior, focusing on the limit of a strong Allee effect. As a preliminary, we  present in Section II a calculation of the MTE based on WKB approximation.
In Section III we add environmental noise to the model: first white noise, and then colored noise.
In Section IV we evaluate the MTE of the population under the simultaneous action of environmental and demographic noises.
Different subsections of Section IV deal with different limits: of short-correlated, long-correlated and (relatively) weak environmental noise.
Section IV also includes a brief discussion of (relatively) strong environmental noise where our results for the MTE coincide with previously known results.
Our main findings are summarized in Section V.

\section{Stochastic population with the Allee effect: preliminaries}
\label{preliminary}
In the absence of environmental noise, a stochastic population exhibiting the Allee effect can be mimicked by three elementary reactions describing binary reproduction, its inverse process and linear decay \cite{AM2010}:
\begin{align}
\label{reactions}
2 A \xra{\lambda} 3 A,&&
3 A \xra{\sigma}  2 A,&&
A \xra{\mu} \varnothing,
\end{align}
with the (constant) reaction rates $\lambda$, $\sigma$ and $\mu$.

\subsection{Deterministic Rate Equation}

\begin{figure}[ht!]
\includegraphics[width=0.35\textwidth]{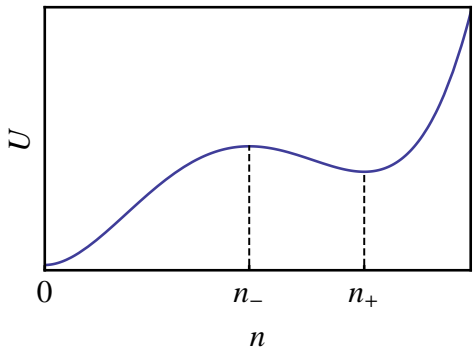}
\caption{Effective potential $U(n) = -\int^{n} f(x)\ d x$ corresponding to Eq.~(\ref{A_MF}) with $\delta^2>0$. Dynamics of the population size according to the mean-field theory corresponds to the coordinate $n(t)$ of an overdamped particle, performing deterministic motion in this potential. Above $n=n_-$ the population grows until it gets established at $n=n_+$, whereas below $n=n_-$ the population goes extinct.}
\label{fig:A_MF}
\end{figure}

The \emph{deterministic} (or mean-field) theory only deals with the mean population size (the number of $A$'s in the system), which is assumed to be large: $n(t)\gg 1$.
The deterministic rate equation has the form
\begin{equation}\label{A_MF}
\dot{n} = f(n)= -\mu n + \frac{\lambda}{2}n^2 - \frac{\sigma}{6}n^3.
\end{equation}
When $\delta^2= 1-8\mu\sigma/(3\lambda^2)>0$, this equation has three fixed points and, therefore, describes a significant Allee effect.
The fixed points $n_0=0$ and $n_+=K(1+\delta)$ are attracting, the fixed point $n_-=K(1-\delta)$ is repelling.
The parameter $K=3\lambda/(2\sigma)$ plays the role of carrying capacity, as it sets the scale of the established population size.
We will assume $K\gg 1$ throughout this work.
Equation (\ref{A_MF}) describes an overdamped dynamics of a classical particle with ``coordinate" $n$ in the effective potential $U(n)=-\int^n f(x)\,d x$, see Fig.~\ref{fig:A_MF}.
The fixed point $n_-$ corresponds to the critical population size for establishment, whereas $n_+$ corresponds to the established population.
That is, according to the mean-field theory, once the initial population size exceeds $n_-$, the population size will approach the fixed point $n_+$ and remain there indefinitely.

\subsection{Stochastic Description and WKB Approximation}\label{sec:NEN}
The mean-field theory, however, disregards fluctuations of the population size around $n=n_+$.
These fluctuations are caused by the demographic noise coming from the discreteness of ``particles"  and from the stochastic character of the reactions (\ref{reactions}).
As $K\gg 1$, these fluctuations are typically small.
However, a rare large fluctuation (an unusual chain of deleterious reactions) ultimately arises and drives the population to extinction.
Indeed, with the death of the last particle, there is no mechanism which would replenish the population.
Because of the Allee effect, it is sufficient for the fluctuation to bring the population below the critical point $n=n_-$, whereupon the population goes extinct essentially deterministically \cite{AM2010}.

Fluctuations of the population size are encoded in the probability $P_n(t)$ to have, at time $t$,  a population of $n$ particles.
The dynamics of $P_n(t)$ is governed by the master equation \cite{vK,Gardiner}
\begin{equation}\label{master}
    \dot{P}_n=\hat{H} P_n = \lambda_{n-1} P_{n-1} +\mu_{n+1} P_{n+1}-(\lambda_n+\mu_n) P_n,
\end{equation}
where
\begin{eqnarray}
\lambda_n=\frac{\lambda n(n-1)}{2}\;\;\text{and}\;\;\mu_n=\frac{\sigma n(n-1)(n-2)}{6}+\mu n \label{rates}
\end{eqnarray}
are the effective birth and death rates.
After a short \emph{relaxation time}, determined by Eq.~(\ref{A_MF}), a long-lived metastable distribution is formed where $P_n(t)$ becomes sharply peaked at $n=n_+$ with an exponential decay towards $n=n_-$ and an almost flat tail at $0<n<n_-$ \cite{AM2010}.
This almost flat tail determines the very slow ``probability leakage" into the absorbing state at $n=0$.
The leakage is described by the (exponentially small) lowest positive eigenvalue $1/\tau$ of the operator $\hat{H}$:
\begin{equation}\label{decay}
    P_n(t)\simeq \pi_n e^{-t/\tau},\;\;n>0.
\end{equation}
Here $\pi_n$ is the lowest excited eigenstate of $\hat{H}$:
 \begin{equation}\label{eigen}
\hat{H} \pi_n=(1/\tau)\,\pi_n
 \end{equation}
which can be identified with the \emph{quasi-stationary distribution} (QSD). In its turn, $\tau$ is equal to the MTE, as
\begin{equation}\label{extinction}
    P_0(t)\simeq 1-e^{-t/\tau},
\end{equation}
see Ref.~\cite{AM2010} for detail.

Employing the large parameter $K\gg1$,  one can accurately calculate the MTE and QSD in this and many  other one-population models \cite{AM2010}.
Our present strategy, however, is different. We will make an additional assumption of a very strong Allee effect where the critical population size $n_-$ is relatively close to the established population size $n_+$ as described by the deterministic system (\ref{A_MF}).
Equivalently, the system (\ref{A_MF}) is close to its saddle-node bifurcation corresponding to the appearance of the fixed points $n_{\pm}$.
In this limit a whole class of population models behaves in a universal way \cite{AM2010}.
Furthermore, in this limit $P_n(t)$ varies with $n$ sufficiently slowly, and the van Kampen system size expansion \cite{vK,Gardiner,Risken} becomes an accurate and controllable procedure.
This procedure replaces the master equation (\ref{master})  by a Fokker-Planck equation, see below.
Then the Langevin equation, equivalent to this Fokker-Planck equation, can be conveniently used for the introduction of environmental noise.

In our example (\ref{reactions}) of three reactions, the strong-Allee-effect regime is achieved when $\delta\ll 1$.
In this case the MTE becomes \cite{AM2010}
\begin{equation}\label{mte2bif}
\tau\simeq\frac{\pi}{\mu\delta}\exp\left(\frac{2}{3}K \delta^3\right).
\end{equation}
The approximate Fokker-Planck equation can be derived from Eq.~(\ref{master}) in a standard manner \cite{vK,Gardiner,Risken}.
It reads
\begin{equation*}
\dot{P}_n = -\frac{d}{dn}\lt[\lt( \lambda_n-\mu_n \rt) P_n \rt] +
              \frac{1}{2} \frac{d^2}{dn^2} \lt[ \lt( \lambda_n+\mu_n \rt) P_n \rt],
\end{equation*}
where $n\gg 1$ is treated as a continuous variable.
Rescaling time, $\bar{t}=\sigma K^2 t/6$, and the population size, $q=n/K$, we obtain
\begin{equation}\label{A_FP_example}
    \dot{\pdf}=\lt\{q \lt[ (q-1)^2-\delta^2 \rt] \pdf \rt\}^\prime +
    \frac{1}{2K} \lt\{ q \lt[ (q+1)^2-\delta^2 \rt] \pdf \rt\} ^{\prime\prime},
\end{equation}
for the continuous probability distribution $\mathcal{P}(q,t)$.
The overbar in $\bar{t}$ is omitted; the primes denote derivatives with respect to $q$.

Since after a short transient $\mathcal{P}(q,t)$ becomes sharply peaked at $q=1+\delta$, and $\delta\ll 1$, we can simplify Eq.~(\ref{A_FP_example}) by putting $q=1$ everywhere except in the combination $q-1$, and by neglecting $\delta^2$ in the second term on the right.
The resulting equation is
\begin{equation}\label{A_FP}
\pder{\pdf}{t} = \lt\{ \lt[ (q-1)^2-\delta^2 \rt] \pdf \rt\}^{\prime}
                +\frac{2}{K}  \pdf^{\prime\prime}.
\end{equation}
This Fokker-Planck equation is equivalent, see \eg\ Ref.~\cite{Gardiner,Risken}, to the following Langevin equation:
\begin{equation}\label{A_SDE}
\dot{q} = -  (q-1)^2+\delta^2
          + \sqrt{\frac{4}{K}}\,\eta_d(t),
\end{equation}
where $\eta_d(t)$ is a white Gaussian noise with zero mean, $\langle\eta_d(t)\rangle = 0$, and $ \langle\eta_d(t_1) \eta_d(t_2) \rangle = \delta(t_1 - t_2) $, whereas the subscript $d$ stands for demographic.
That is, close to the saddle-node bifurcation, the demographic noise is effectively Gaussian, white (that is, uncorrelated) and additive.
From now on, when discussing the correlation properties of a noise, we will always mean the environmental noise.

Equations (\ref{A_FP}) and (\ref{A_SDE}) hold (up to rescaling, and close to the saddle-node bifurcation) for a whole class of single-population models which exhibit the Allee effect.
Furthermore, these equations represent a truly paradigmatic model of escape from the vicinity of an attracting fixed point due to a weak additive white Gaussian noise.
This model has appeared in numerous contexts, and the mean time to escape in this model is well known \cite{vK,Gardiner,Risken}.
We will proceed, however, as if we were unaware of these classical results.
This is because we want, as a preliminary for the following material, to briefly outline how to evaluate the mean time to escape by using the weak-noise WKB approximation due to Freidlin and Wentzell \cite{FW98}, see also Refs.~\cite{Dykman,Graham}.
As we will see shortly, this approximation is readily extendable to the situation of our interest where weak demographic and environmental noises are both present.

\begin{figure}[ht!]
\includegraphics[width=0.4\textwidth]{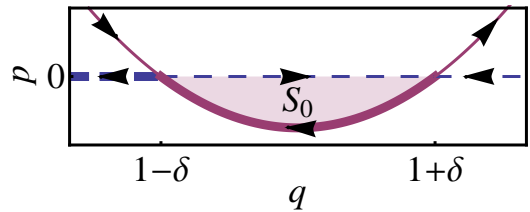}
\caption{(Color online)
Zero-energy phase trajectories of the Hamiltonian (\ref{A_H}).
The dashed line shows the  $p=0$ trajectory.
The solid curve depicts $p_0(q)$ from Eq.~(\ref{sep1}).
The area of the shaded region is equal to $S_0$ from Eq.~(\ref{S0}).
The thick line shows the optimal (most likely) path to extinction. The mean time to extinction can be estimated as $\tau \sim \exp(KS_0)$, see Eq.~(\ref{MTE0}).
}
\label{fig:A_traj}
\end{figure}

WKB approximation predicts the mean time to escape which coincides with the MTE.
We look for the solution of Eq.~(\ref{A_FP}) as $P(q,t)=\pi(q) \exp(-t/\tau)$ and assume that $\tau$ is exponentially large with respect to the parameter $K$, see Eq.~(\ref{mte2bif}).
This justifies a quasi-stationary formulation for $\pi(q)$.
Then the WKB ansatz $\pi(q)=\exp[K S(q)]$ yields, in the leading order in $1/K$, a stationary Hamilton-Jacobi equation $H(q,\partial_q S) \simeq 0$, where
\begin{equation}\label{A_H}
H(q,p) = 2p^2- \lt[ (q - 1)^2 - \delta^2 \rt] p,
\end{equation}
and $p$ is the ``momentum" canonically conjugate to the ``coordinate" $q$.
We should only deal with zero-energy trajectories.
One type of zero-energy trajectories also have a zero momentum, $p=0$.
For $p=0$ the Hamilton equations read $\dot{q}=-(q - 1)^2 + \delta^2,\; \dot{p}=0$, so these are (deterministic) \emph{relaxation trajectories}.
The escape is encoded by an \emph{activation trajectory}: a zero-energy trajectory with $p\neq 0$.
This trajectory,
\begin{equation}\label{sep1}
p = p_0(q)=(1/2) \left[(q-1)^2-\delta^2\right]
\end{equation}
is an instanton or, in a more mathematical language, a heteroclinic connection which exits the fixed point $q=1+\delta,\,p=0$ and enters the fixed point $q=1-\delta,\,p=0$.
Then the population size flows toward $q=0$ along a deterministic trajectory which does not cost action, see Fig.~\ref{fig:A_traj}.
(The latter segment corresponds to the aforementioned almost flat tail of the probability distribution.)
Therefore, in the leading WKB order, we only need to calculate the mechanical action along the the activation trajectory from $q=1+\delta$ to $q=1-\delta$:
\begin{equation}
\label{S0}
    S_0=\int_{1+\delta}^{1-\delta} p_0(q)\,dq =  \frac{2}{3}\,\delta^3.
\end{equation}
As a result, the MTE due to the demographic noise is, up to a pre-exponent \cite{AM2010,EK}
\begin{equation}\label{MTE0}
    \tau_0 \sim \exp\left(\frac{2}{3}\,K \delta^3\right).
\end{equation}
This result (which is of course well known)  agrees with the more accurate result presented in Eq.~(\ref{mte2bif}).
The quantity $(2/3)K\delta^3$ is nothing but the Arrhenius factor $\Delta U_0/\Theta$, where $\Theta=2/K$ is the effective temperature [see Eqs.~(\ref{A_FP}) and (\ref{A_SDE})], $\Delta U_0=U_0(q=1-\delta)-U_0(q=1+\delta)$ is the potential barrier height, and $U_0(q)=(1/3)(q-1)^3-\delta^2 q$ is the potential corresponding to the force $f_0(q)=-(q-1)^2+\delta^2$.

Now we can see that the large parameter of the WKB theory is $K\delta^3 \gg 1$.
Actually, this could have been seen directly from Eq.~(\ref{A_FP}):  by rescaling $(q-1)/\delta \to q$ and $\delta t\to t$, one obtains the equation
\begin{equation}\label{A_FP_resc}
\pder{\pdf}{t} = \left[(q^2-1) \pdf \right]^{\prime}
                +\frac{2}{K\delta^3}  \pdf^{\prime\prime}
\end{equation}
containing a single parameter $K\delta^3$.
When this parameter is large, the quasi-stationary distribution is sharply peaked around the attracting fixed point $q=1$, thus validating the WKB approximation.

\section{Incorporating Environmental Noise}
Environmental variations affect the population dynamics by modulating the birth and death rates.
As a result, the bifurcation parameter $\delta^2$, which enters Eq.~(\ref{A_SDE}),  becomes time-dependent: $\delta^2(t) = \delta_0^2 - \xi(t)$, where the zero-mean random process $\xi(t)$ is independent of the demographic noise $\eta_d(t)$.
Now Eq.~(\ref{A_SDE}) becomes
\begin{equation}\label{G_SDE}
\dot{q} = -  (q-1)^2+\delta_0^2 - \xi(t) + \sqrt{\frac{4}{K}}\,\eta_d(t).
\end{equation}
In fact, making the reaction rates noisy will in general affect not only $\delta^2$ but also $K$.
However, if the environmental noise is sufficiently weak,  an account of this effect only leads to a subleading correction which we will ignore.

In contrast to the Langevin equation which describes a combined action of the demographic and environmental noise in the absence of the Allee effect \cite{OM}, the demographic and environmental noises in Eq.~(\ref{G_SDE}) are both additive.

\subsection{White Noise}
\label{sec:White Noise}
The nature of environmental noise manifests itself in the properties of the random process $\xi(t)$.
The simplest environmental noise to consider is a white noise of a given intensity $D$: $\xi(t) = \sqrt{2 D}\, \eta_e(t)$, with $\eta_e(t)$ having the same properties as $\eta_d(t)$, the two being independent.
Now Eq.~(\ref{G_SDE}) reads
\begin{equation}\label{B_SDE_separated}
\dot{q} = -  (q - 1)^2 + \delta_0^2
+ \sqrt{\frac{4}{K}}\, \eta_d(t)
+ \sqrt{    2 D    } \,\eta_e(t).
\end{equation}
As $\eta_d$ and $\eta_e$ are statistically independent Gaussian processes, their weighted sum is another Gaussian, and we obtain
\begin{equation}\label{B_SDE_combined}
\dot{q} = - (q - 1)^2 + \delta_0^2
+\left(\frac{4}{K} + 2 D\right)^{1/2} \eta_t(t),
\end{equation}
where $\eta_t(t)$ is a white Gaussian noise with the same properties as $\eta_d(t)$ and $\eta_e(t)$.
Equation (\ref{B_SDE_combined}) coincides with Eq.~(\ref{A_SDE}), except for a greater noise intensity.
If the combined noise is still sufficiently weak, we can again use WKB approximation, or simply replace $K$ in Eq.~(\ref{MTE0}) by $K/(I+1)$, where
\begin{equation}\label{I_def}
I= \frac{1}{2} D K
\end{equation}
is the ratio of the environmental and demographic noise intensities.
This corresponds to the MTE
\begin{equation}\label{B_S}
\tau_{\text{white}} \sim e^{KS_{\text{white}}}\;\;\;\text{with}\;\;\;S_{\text{white}}=\frac{2 \delta_0^3}{3(I+1)}.
\end{equation}
The reduction of the purely demographic action $S_0$ by a factor of $I+1$ has a great impact on the MTE, especially if $I$ is large.
When $I\gg 1$, $S_{\text{white}}$ scales as $1/I$.
As a result, the demographic parameter $K$ cancels out in the expression for the MTE, and $\tau \sim \exp[(4\delta_0^3) /(3D)]$ is dominated by the environmental noise.

\subsection{Colored Noise}
Let us now return to Eq.~(\ref{G_SDE}) and choose $\xi(t)$ to be a colored (positively correlated) Gaussian noise, as modeled by the Ornstein-Uhlenbeck random process \cite{vK,Gardiner,Risken,Luszka}.
The autocorrelation function of the Ornstein-Uhlenbeck random process is $\langle \xi(t_1) \xi(t_2)\rangle = (D/\tau_c) \exp(-|t_1-t_2|/\tau_c)$, where $\tau_c$ is the correlation time of the noise, and $D$ is the noise intensity.
As Eq.~(\ref{G_SDE}) is written in rescaled variables, $\tau_c$ and $D$ are also assumed to be properly rescaled.
The variance of the  Ornstein-Uhlenbeck noise is equal to $D/\tau_c$.
The Ornstein-Uhlenbeck process can be conveniently represented by a Langevin equation,
\begin{equation}\label{OU}
\dot{\xi} = - \frac{\xi}{\tau_c} + \frac{\sqrt{2 D}}{\tau_c}\,\eta_e(t).
\end{equation}
When $\tau_c$ tends to zero, Eq.~(\ref{OU}) turns into $\xi(t)=\sqrt{2 D} \,\eta_e(t)$, and the white-noise limit is recovered.
For finite $\tau_c$ we have to deal with two coupled scalar Langevin equations (\ref{G_SDE}) and (\ref{OU}) with mutually independent white Gaussian noises $\eta_e(t)$ and $\eta_d(t)$.
These Langevin equations are equivalent, see \eg\ \cite{Gardiner,Risken}, to a Fokker-Planck equation for the joint probability distribution $\pdf(q,\xi,t)$ of the rescaled population sizes $q$ and the environmental fluctuations $\xi$:
\begin{align}\label{C_FP}
\pder{\pdf}{t}=
                   \pder{}      {q}     \lt\{ \lt[ (q - 1)^2 - \delta_0^2 \rt] \pdf \rt\} +
\xi                \pder{\pdf}  {q}      +
\frac{2}{K}        \pder{^2\pdf}{q^2}    + \notag\\
\frac{1}{\tau_c}   \pder{}      {\xi}   (\xi\pdf) +
\frac{D}{\tau_c^2} \pder{^2\pdf}{\xi^2}.
\end{align}

\section{WKB Analysis}
Multi-dimensional Fokker-Planck equations, like Eq.~(\ref{C_FP}), are in general hard to solve. A plethora of
approximate methods of their solution have been developed in the literature; the reader is referred to Refs. \cite{Gardiner,Risken,Luszka} for their description.  Among these approximate methods, the WKB formalism is
especially suitable for the analysis of rare, noise-induced transitions which arise when the noise is typically
small.   Correspondingly, we assume throughout this paper that both environmental and demographic noises are sufficiently weak (we will obtain the corresponding criteria \emph{a posteriori}).
Mathematically, this means that the coefficients of the two diffusion terms in Eq.~(\ref{C_FP}) are small.
In this case the time history of $\pdf(q,\xi,t)$, as described by Eq.~(\ref{C_FP}), is the following.
After a relatively short transient (typically of duration $\sim 1/\delta_0$), $\pdf(q,\xi,t)$ develops a sharp peak at $q=1+\delta_0,\, \xi=0$, as predicted by Eq.~(\ref{C_FP}) with $K\to\infty$ and $D\to 0$.
The small diffusion terms arrest the peak growth so quasi-stationarity sets in.
At the same time, the probability very slowly leaks toward the point $q=1-\delta_0,\, \xi=0$ beyond which the distribution is almost flat, which corresponds to a quick escape toward the absorbing state at $q=0$.
This late-time dynamics is described by the eigenstate of the Fokker-Planck operator with the (exponentially small) lowest positive eigenvalue.
Combining this knowledge with the WKB (or Freidlin-Wentzel) ansatz for the quasi-stationary distribution, we can write
\begin{align}
\label{C_pdf_QSD}
\pdf(q,\xi,t)=\pi(q,\xi)\,e^{-\frac{t}{\tau}}, && \pi(q,\xi)=e^{-K S(q,\xi)}.
\end{align}
Then Eq.~(\ref{C_FP}) yields, in the leading order in $1/K$, a Hamilton-Jacobi equation with two degrees of freedom: $H(q,\xi,\partial_q S, \partial_{\xi} S) \simeq 0$, with effective Hamiltonian
\begin{equation}\label{C_H0}
H = 2 p^2 - \lt[ (q - 1)^2 - \delta_0^2 \rt] p  - \xi p  + \frac{2 I}{\tau_c^2} P^2 - \frac{1}{\tau_c} \xi P.
\end{equation}
The two  momenta $p$ and $P$ are conjugate to the ``coordinates" $q$ and $\xi$, respectively. 
The Hamilton equations are
\begin{align*}
\dot{q} &= \delta_0^2-\xi -(q - 1)^2 + 4 p, & \dot{\xi} &= - \frac{1}{\tau_c} \xi + \frac{4I}{\tau_c^2} P, \notag\\
\dot{p} &=  2 (q-1) p,                                    & \dot{P  } &=   \frac{1}{\tau_c} P   + p.
\end{align*}
Now we should look for an instanton: a zero-energy but non-zero momentum trajectory which exits, at $t=-\infty$, the fixed point $(q,\xi,p,P)=(1+\delta_0,0,0,0)$ of this Hamiltonian flow, and enters, at $t=\infty$, the fixed point $(q,\xi,p,P)=(1-\delta_0,0,0,0)$.
Once the instanton is found, we can compute the action along it,
\begin{equation}\label{H_S}
S =S_q+S_{\xi}= \int_{-\infty}^{\infty} \lt( p \dot q + P \dot \xi \rt) dt,
\end{equation}
and evaluate the MTE from
\begin{equation}\label{MTE}
\tau \sim e^{K S}.
\end{equation}
Let us apply one more rescaling transformation:
\begin{align}\label{C_trans}
&
\begin{aligned}
\bar{q} &= (q-1) / \delta_0,  \\
\bar{p} &=  p    / \delta_0^2,
\end{aligned}
&&
\begin{aligned}
\bar{\xi} &= \xi  / \delta_0^2, \\
\bar{P  } &= P    / \delta_0,
\end{aligned}
&&
\bar{t}= \delta_0\,t.
\end{align}
Omitting the overbars in the equations of motion, we obtain
\begin{align}
\dot{q}   &= 1- \xi - q^2 + 4 p,                   \label{C_HE_q} \\
\dot{p}   &=            2 q p,                     \label{C_HE_p} \\
\dot{\xi} &= - \frac{1}{T} \xi + \frac{4I}{T^2} P, \label{C_HE_xi} \\
\dot{P  } &=   \frac{1}{T} P   + p.                \label{C_HE_P}
\end{align}
As we can see, the dynamics is controlled by two dimensionless parameters: $I=DK/2$ and $T=\delta_0 \tau_c$, the latter being the ratio of the correlation time of the environmental noise and the relaxation time of the system without noise.
As we will see, sometimes it is more convenient to use the rescaled variance $V=I/T$ of the environmental noise instead of the rescaled intensity $I$.

Equations (\ref{C_HE_q})-(\ref{C_HE_P}) are Hamiltonian, as they stem from the Hamiltonian
\begin{equation}\label{C_H}
\bar{H}  =   2 p^2-  \lt( q^2 - 1 \rt) p   - \xi p  +  \frac{2I}{T^2} P^2 -  \frac{1}{T} \xi P.
\end{equation}
In the rescaled variables, the instanton connects the fixed points $(1,0,0,0)$ and $(-1,0,0,0)$.
Denoting the action along this instanton by $\bar{S}$, we can express the original action $S$ which appears in Eq.~(\ref{H_S}) as $S=\delta_0^3 \bar{S}$.
As a result, Eq.~(\ref{MTE}) becomes
\begin{equation}\label{SSS}
   \tau\sim e^{K \delta_0^3 \bar{S}}.
\end{equation}

The two-dimensional Hamiltonian system (\ref{C_H}) is in general non-integrable, as the only available integral of motion -- the Hamiltonian itself -- is insufficient for integrability.
As a result, it is impossible to find the instanton analytically for arbitrary $I$ and $T$.
Perturbative solutions of different types are possible, however, in several regions of the $(I,T)$ plane; seeking such solutions will be our main strategy for the remainder of the paper.
We will refer to the limits of small and large $T$ (with the criteria derived \textit{a posteriori}) as the short- and long-correlated (environmental) noise, respectively.
The limits of small and large $I$ will be called the weak and strong (environmental) noise,  respectively.

\subsection{Short-Correlated Noise}

\subsubsection{Leading order in $T\ll 1$}

For very small $T$ the right-hand-sides of Eqs.~(\ref{C_HE_xi}) and (\ref{C_HE_P}) include large factors.
As a result, $\xi$ and $P$  quickly adjust to the current value of $p(t)$ which slowly evolves with time: $\xi(t)\simeq -4Ip(t)$ and $P(t)\simeq-Tp(t)$.
Plugging this $\xi(t)$ in Eq.~(\ref{C_HE_q}), we obtain
\begin{equation*}
    \dot{q}=1-q^2+4(I+1)p.
\end{equation*}
This equation and Eq.~(\ref{C_HE_p}) are Hamilton equations, with the effective Hamiltonian
\begin{equation*}
    H_{1}(q,p)=2 (I+1) p^2+(1-q^2)p
\end{equation*}
which coincides, up to rescaling, with Eq.~(\ref{A_H}).  The escape instanton satisfies
\begin{equation*}
    p_0(q)=\frac{q^2-1}{2(I+1)},
\end{equation*}
so the rescaled action in the leading order is
\begin{equation}
    \bar{S}_0=\int_{1}^{-1} p_0(q) dq=\frac{2}{3(I+1)}
    \label{barS0}
\end{equation}
which, along with Eq.~(\ref{SSS}), yields Eq.~(\ref{B_S}) for the MTE. One can also obtain explicit solutions for $q(t)$ and $p(t)$ \cite{shift} in the leading order in $T\ll 1$:
\begin{equation}
\label{D_qp0}
q_0(t) = -\tanh t,\;\;\;\;\; p_0(t) = -\frac{1}{2(I+1) \cosh^2 t}.
\end{equation}
Correspondingly,
\begin{equation}
\label{D_xiP0}
\xi_0(t) = \frac{2I}{(I+1) \cosh^2 t},\;\;P_0(t) = \frac{T}{2(I+1) \cosh^2 t},
\end{equation}
where the subscript $0$ stands for the leading-order quantities.
As one can see, $q_0(t)$ does not depend on $I$, whereas the magnitude of the environmental fluctuation $\xi_0(t)$ goes up with $I$ and then saturates: $\xi_0(t, I\gg 1)=2 \cosh^{-2} t$.
The magnitudes of the momenta $p$ and $P$ go down as $I$ increases.
This is expected on physical grounds: the stronger is the environmental noise, the smaller are the momenta needed for escape, leading to a smaller action and a shorter escape time.
As both $P_0(t)$ and $\xi_0(t)$ are even functions of time, the environmental noise does not contribute to the action in the first order of $T$.

\begin{figure}[ht!]
\includegraphics[width=0.5\textwidth]{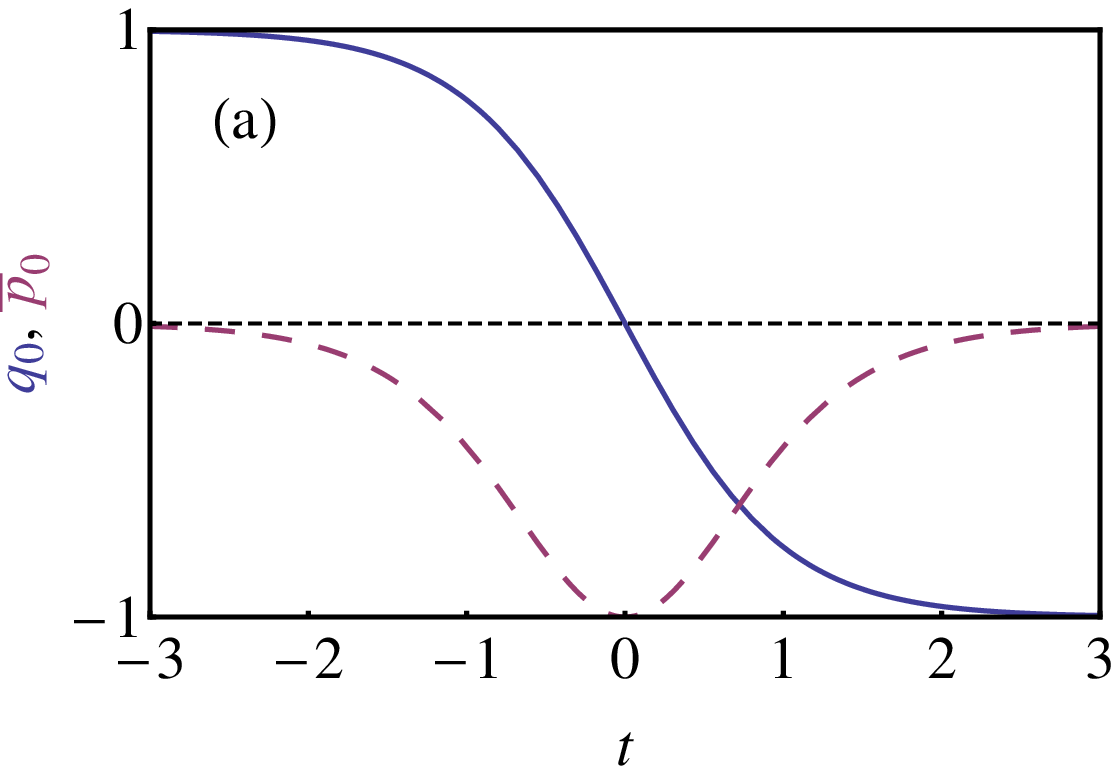}
\includegraphics[width=0.5\textwidth]{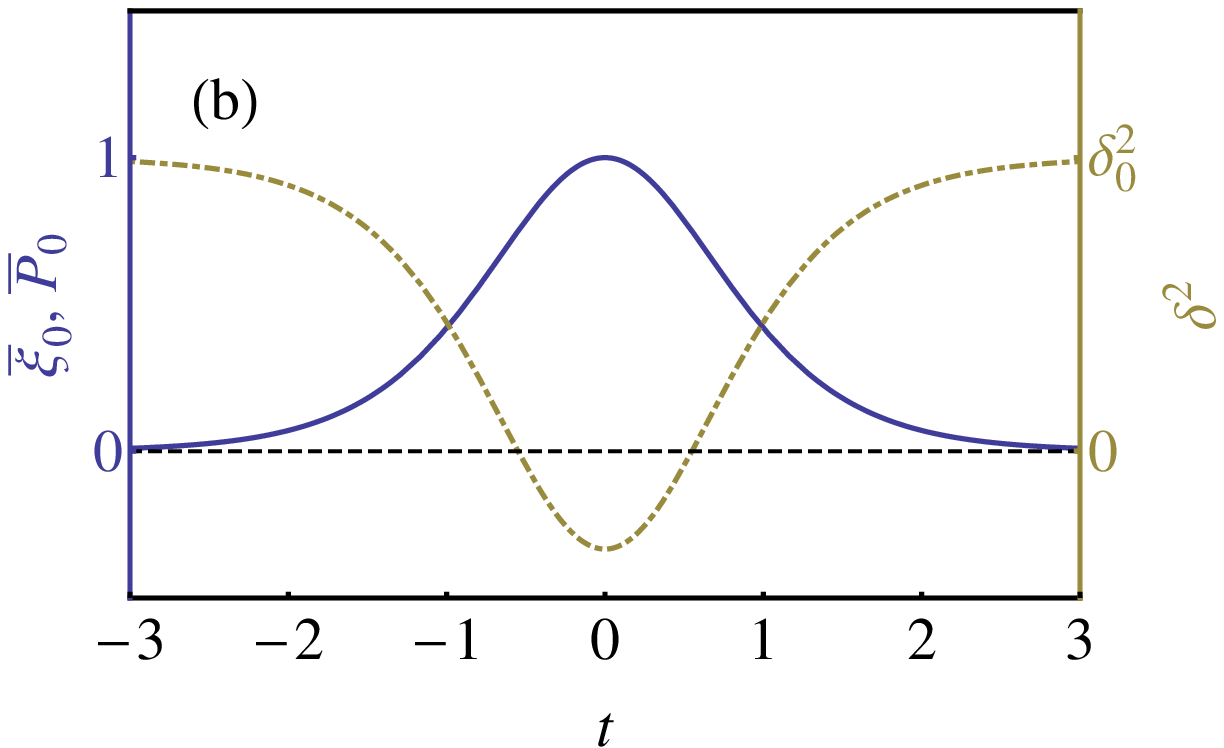}
\caption{(Color online)
The path to extinction for short-correlated noise. Shown is the optimal path (a) and the optimal environmental fluctuation (b).
The values of the functions are normalized by their extrema (and denoted by overbars). All of the functions vary over the same time scale.
The population size $q_0(t)$ and the (normalized) conjugate momentum $\bar{p}_0(t)=2(I+1) p_0(t)$ are depicted by the solid and dashed curves, respectively, in (a). The population transition from the metastable state, $q_0=1$, to the verge of extinction, $q_0=-1$, is evident.
The optimal environmental fluctuation $\xi_0(t)$ and its conjugate momentum $P_0(t)$ coincide after normalization, so the solid line in (b) represents both $\bar{\xi}_0(t)=(1/2)(1+1/I)\,\xi_0(t)$ and $\bar{P}_0(t)=(2/T) (I+1)\,P_0(t)$. The time-dependent bifurcation parameter $\delta^2(t)$ is depicted by the dot-dashed curve in (b).}
\label{fig:D_I}
\end{figure}

Figure \ref{fig:D_I} depicts $q_0, p_0, \xi_0$ and $P_0$ versus time, along with $\delta^2(t)$.
Note that, for $I>1$,  the effective time-dependent bifurcation parameter $\delta^2(t)=\delta_0^2(1-\xi_0(t))$ becomes negative on a time interval around $t=0$.
This change of sign, however, occurs on the same time scale as that of the $q$- and $p$-dynamics, and does not lead to any qualitative change in the character of solution \cite{different}.

\subsubsection{Subleading order in $T\ll 1$}
Now we calculate the next-order correction to the white-noise result.
Using $T$ as a small parameter, we look for the solutions of Eqs.~(\ref{C_HE_xi}) and (\ref{C_HE_P}) as $\xi(t)=\xi_0(t)+ T^2 \xi_2(t)+\dots$ and $P(t)=T P_1(t)+T^2 P_2(t)+\dots$ (the odd powers of $T$ in the expansion of $\xi$ turn out to be absent).
We substitute these expressions into Eqs.~(\ref{C_HE_xi}) and (\ref{C_HE_P}) and demand cancellation in every order in $T$.
This procedure yields  $\xi$ and $P$ expressed via $p(t)$:
\begin{eqnarray}
\xi(t) &=& -4 I \left[p(t)+T^2 \ddot{p}(t)+T^4  p^{(4)}(t) +\dots\right], \label{xi}\\
  P(t) &=& -  T \left[p(t)+T    \dot{p}(t)+T^2 \ddot{p}(t) +\dots\right], \label{P}
\end{eqnarray}
where the leading-order terms coincide with those we obtained previously.
Combining Eq.~(\ref{xi}) with Eq.~(\ref{C_HE_q}), we obtain in the leading and subleading orders
\begin{equation}
\label{D_HE_qp}
\dot{q}= 1- q^2+ 4(I+1)(p + \eps \ddot{p}),
\end{equation}
where $\eps = I T^2 / (I+1)\ll 1$.
Equations (\ref{C_HE_p}) and (\ref{D_HE_qp}) make a closed set and can be solved perturbatively in $\eps$, by setting $q(t)=q_0(t)+\eps q_1(t)$, $p(t)=p_0(t)+\eps p_1(t)$ with $q_0$ and $p_0$ from Eq.~(\ref{D_qp0}).
Eliminating $p_1$ we obtain, in the first order in $\eps$:
\begin{equation*}
\ddot{q_1} + \lt( \frac{6}{\cosh^2 t}-4 \rt) q_1=-\frac{24 \sinh t}{\cosh^5 t}.
\end{equation*}
We are looking for the forced solution of this linear equation which obeys zero boundary conditions at $t\to \pm \infty$.
This solution turns out to be elementary:
\begin{equation}\label{q1}
q_1(t) =\frac{4\sinh t}{\cosh^3 t}.
\end{equation}
The corresponding forced solution for $p_1$ which vanishes at $t\to \pm \infty$ is
\begin{equation}\label{p1}
p_1(t) = -\frac{2 \sinh^2 t}{(I+1)\cosh^4 t}.
\end{equation}
Now we can calculate $S_q$: the contribution of the $(q,p)$ subsystem to the action (\ref{H_S}):
\begin{align*}
S_q & =
\int_{-\infty}^\infty \lt[ p_0 \dot{q}_0 +
              \eps (p_0 \dot{q}_1 +
                           p_1 \dot{q}_0) \rt] dt = \notag\\
& = \frac{2}{3(I+1)} -
    \frac{8IT^2}{15(I+1)^2}+{\cal O}(T^3).
\end{align*}

Once $p(t)$ is found up to the second order in $\eps$, the sub-leading corrections for $\xi$ and $P$ can be calculated from Eqs.~(\ref{xi}) and (\ref{P}), respectively.
We skip these formulas here and focus on calculating the important correction to the action coming from the $(\xi,P)$ subsystem.
Using  Eqs.~(\ref{xi}) and (\ref{P}), we obtain
\begin{eqnarray}
  S_\xi &=& \int_{-\infty}^\infty P \dot{\xi} dt
     =4 I T^2 \int_{-\infty}^\infty \dot{p_0}^2 dt + {\cal O}(T^3) \nonumber \\
 &=& \frac{16 IT^2}{15 (I+1)^2}+{\cal O}(T^3). \label{Sxiw}
\end{eqnarray}
The total action is then
\begin{equation}\label{D_S}
\bar{S}= S_q + S_{\xi} \simeq \bar{S}_{\text{white}} \left[1+\frac{4IT^2}{5(I+1)} \right]
\end{equation}
where $\bar{S}_{\text{white}} = (2/3) (I+1)^{-1}$.
That is, for an almost white noise, $T\ll 1$, the MTE (\ref{SSS}) is longer, and so the extinction risk is lower, than for the white noise of the same \emph{intensity} $I$.
However, if we keep the \emph{variance} constant, $V=I/T=\text{const}$, then $T>0$ reduces the MTE and increases the extinction risk, as follows from the leading-order result (\ref{barS0}): $\bar{S}_0=(2/3)(1+VT)^{-1}\simeq (2/3)(1-VT)$.

\begin{figure}[ht!]
\includegraphics[width=0.45\textwidth]{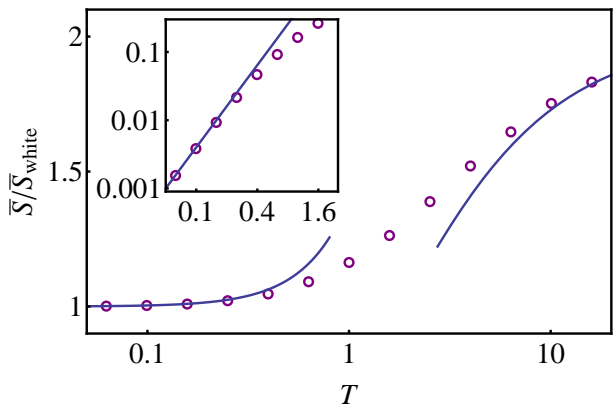}
\caption{(Color online)
Analytic (lines) and numerical (circles) results for $\bar{S}/\bar{S}_{\text{white}}$ versus $T$ on a  semi-logarithmic scale for $I=1$. Left line: short-correlated noise theory, Eq.~(\ref{D_S}).
Right line: long-correlated noise theory, Eq.~(\ref{E_S}).
The inset shows,  on a log-log scale, the small-$T$ correction $\bar{S}/\bar{S}_{\text{white}}-1$, see Eq.~(\ref{D_S}). The asymptote's slope is $2$, so
the $T^2$ behavior of the correction is evident.
One can see that, when the effects of two noises with the same \emph{intensity} are compared, the noise with the shorter correlation time demands a smaller action, thereby causing a quicker extinction. For two noises with the same \emph{variance} (not shown), the action goes down as $T$ increases, so the shorter-correlated noise is less dangerous extinction-wise in this case.}
\label{fig:DE_T_sim}
\end{figure}

Figure \ref{fig:DE_T_sim} shows a comparison of the action from Eq.~(\ref{D_S}) with the results of our numerical calculations for $I=1$ and  different $T$.
The numerical results were obtained by computing the instanton solution of  the full set of equations (\ref{C_HE_q})-(\ref{C_HE_P}) by a shooting method, and then evaluating the action integral in Eq.~(\ref{H_S}) numerically, see Ref.~\cite{KamMe} for details.

\subsection{Long-Correlated Noise}
Here analytic progress is possible due to time-scale separation.
Indeed, for sufficiently large $T$ (we will obtain the criterion \textit{a posteriori}) the right-hand-side of Eq.~(\ref{C_HE_xi}) is small.
Therefore, $\xi(t)$ varies slowly (adiabatically) compared with $q(t)$ and $p(t)$, and can be treated as constant when dealing with the fast $(q,p)$ sub-system.
Equations (\ref{C_HE_q}) and (\ref{C_HE_p}) with $\xi=\text{const}$ are Hamilton equations with the Hamiltonian
\begin{equation}\label{E_H}
H_0 = 2 p^2 - \lt( q^2-\nu^2 \rt) p,
\end{equation}
where we have defined $\nu^2=1-\xi$.
This Hamiltonian coincides with that of Eq.~(\ref{A_H}) up to rescaling.
We immediately obtain the instanton solution in parametric form:
\begin{equation}\label{E_arc}
p=p(q) = \tfrac{1}{2} \lt( q^2-\nu^2 \rt).
\end{equation}
The time-dependent solutions are
\begin{equation}
\begin{aligned}\label{E_qp}
& q(t) = -       \nu             \tanh       \nu t, \\
& p(t) = - \frac{\nu^2}{2} \frac{1} {\cosh^2 \nu t}.
\end{aligned}
\end{equation}
The characteristic fast time scale is $1/\nu$, with a yet unknown $\nu$.
We assume here (and will check \textit{a posteriori}) that $\nu^2=1-\xi>0$.

\begin{figure}[ht!]
\includegraphics[width=0.4\textwidth]{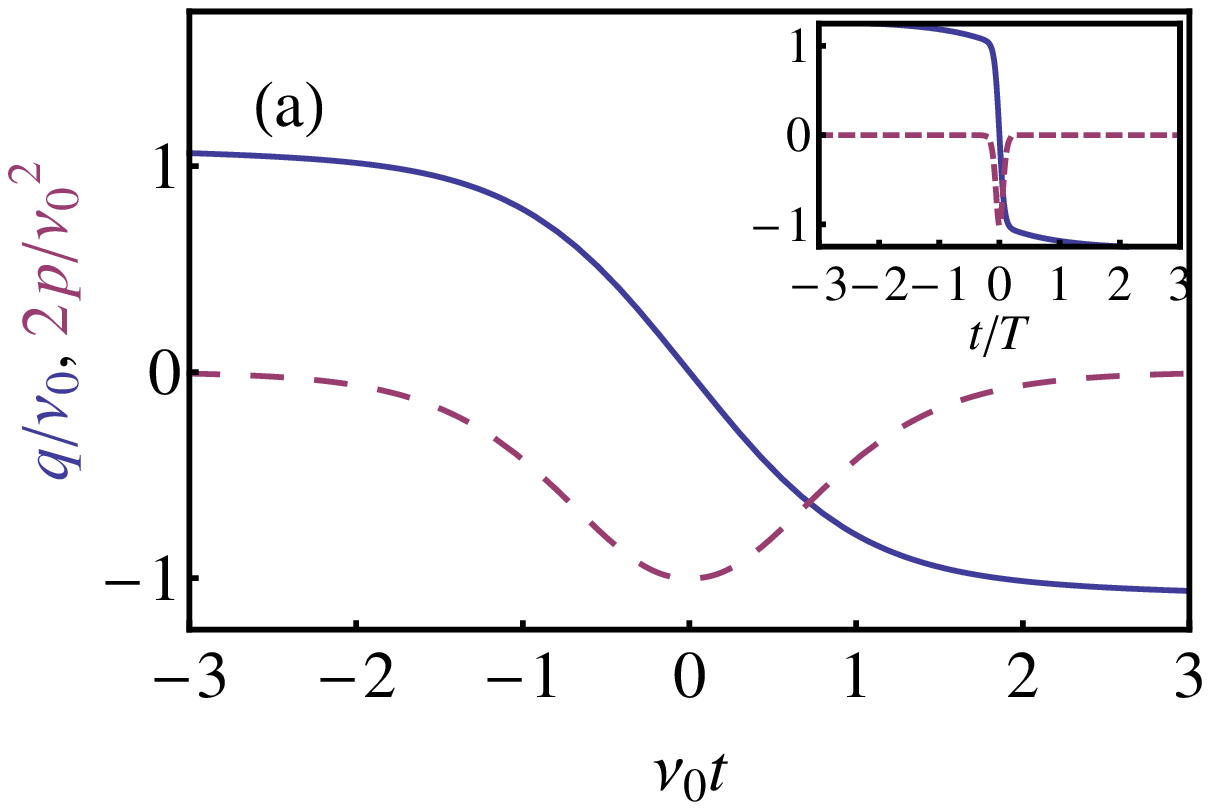}
\includegraphics[width=0.4\textwidth]{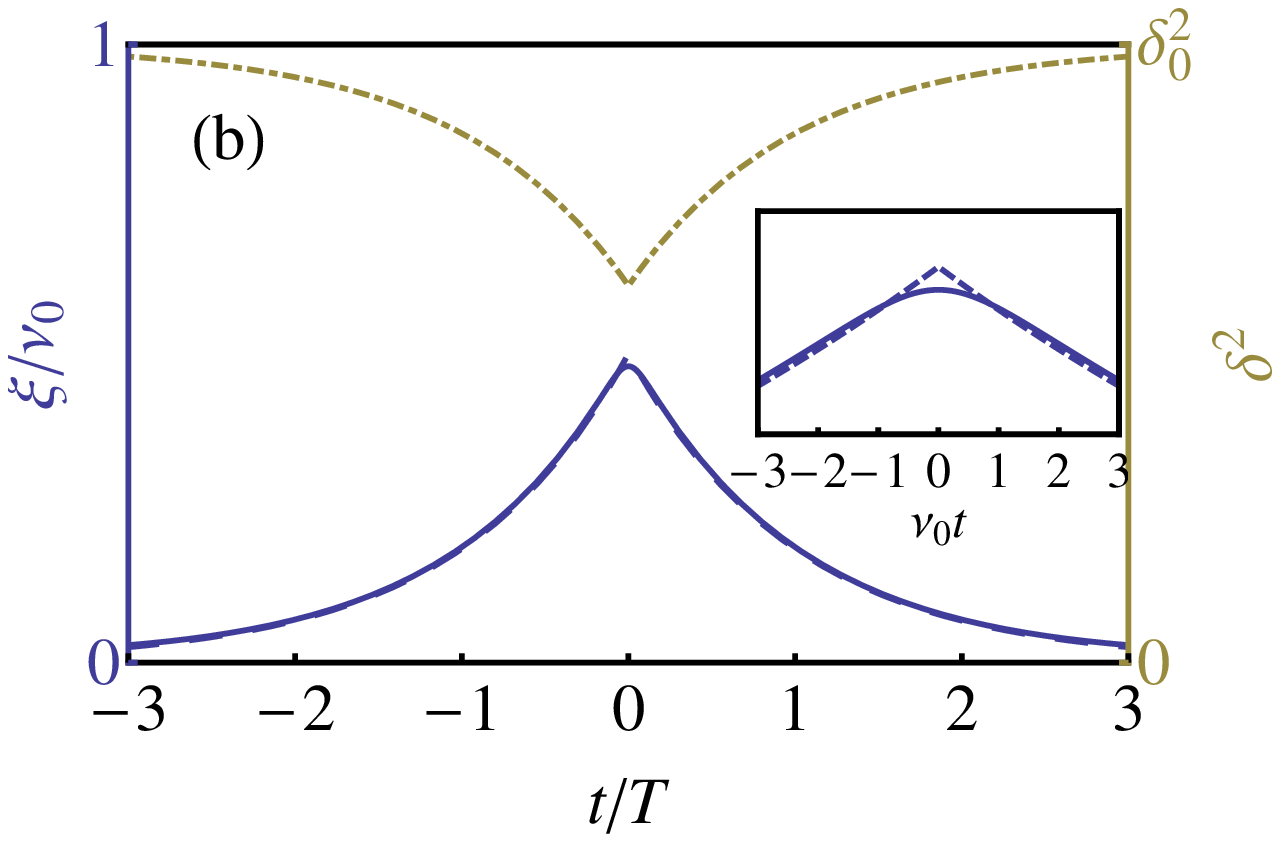}
\includegraphics[width=0.4\textwidth]{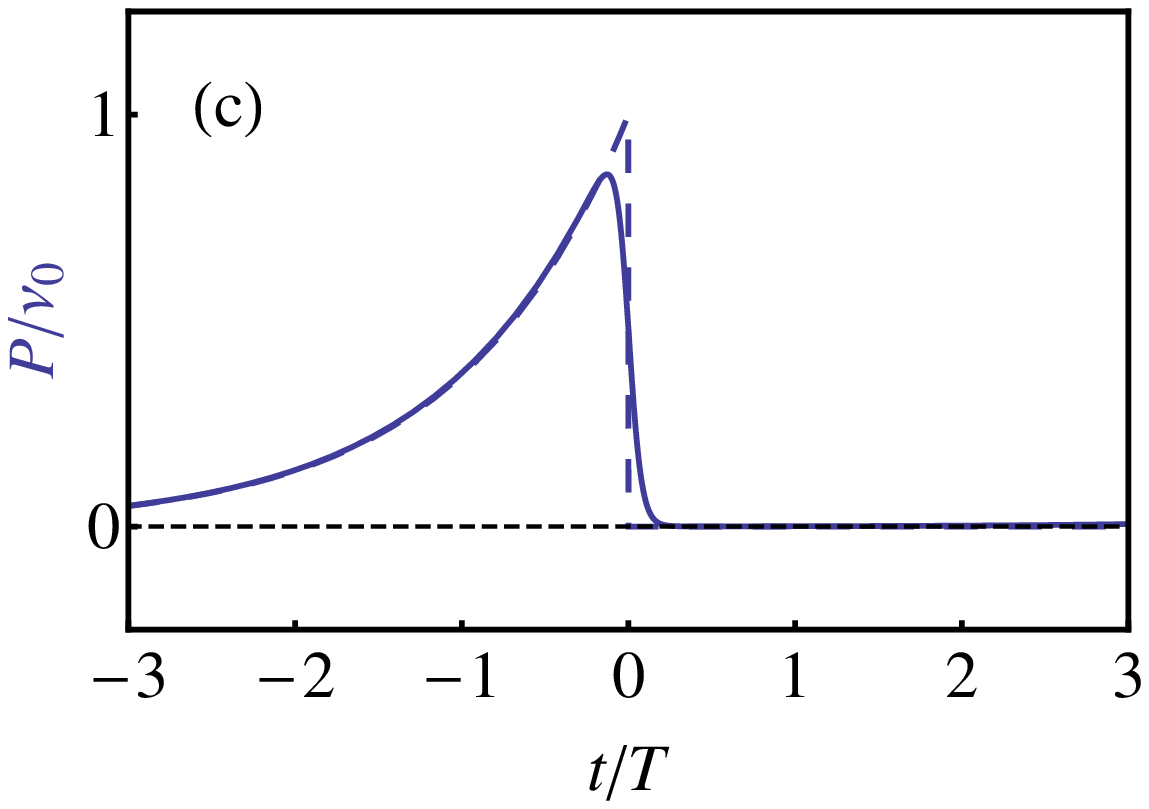}
\caption{(Color online)
The path to extinction for long-correlated noise. Shown are the optimal path (a) and optimal environmental fluctuation (b,c), with $T=16$ and $I=4$ (so $\nu_0 \simeq 0.78$).
(a) Analytic results for $q/\nu_0$ (solid line) and $p/\nu_0^2$ (dashed line)
versus the fast time $\nu_0 t$ (main panel) and slow time $t/T$ (inset).
Numerical results are not shown as they are indistinguishable from the analytic results.
(b) Numerical (solid line) and analytic (dashed line) results for $\xi/\nu_0$
versus the slow (main panel) and fast (inset) times.
Also shown is the time-dependent bifurcation parameter $\delta^2(t/T)$ predicted analytically.
(c) Numerical (solid line) and analytic (dashed line) results for $P(t/T)/\nu_0$. The population transition from the metastable state, $q_0=1$, to the verge of extinction, $q_0=-1$, is evident. Here it happens on a time scale much shorter than the time scale of environment variation [compare the inset in (a) with the main panel in (b)].}
\label{fig:E_I}
\end{figure}

Now we turn to the slow sub-system $(\xi,P)$.
Differentiating Eq.~(\ref{C_HE_xi}) with respect to time and using Eq.~(\ref{C_HE_P}), we obtain an exact linear second-order equation for $\xi(t)$:
\begin{equation}\label{E_xi}
\ddot{\xi}(t) -\frac{\xi(t)}{T^2} = \frac{4I}{T^2}\,p(t),
\end{equation}
which has to be solved with the boundary conditions $\xi(t\to\pm \infty)=0$.
In our adiabatic approximation $p(t)$, entering the forcing term, is given by Eq.~(\ref{E_qp}), but  $\nu=\nu(t)=[1-\xi(t)]^{1/2}$ is now time-dependent.
However, the time scale $T$, determined by the left-hand side of Eq.~(\ref{E_xi}), is supposedly much longer than the time scale of the forcing.
Therefore, the forcing pulse, see Eq.~(\ref{E_qp}) for $p$, can be approximated by a delta-function with the proper amplitude:
\begin{equation}
\label{E_xi_delta}
\ddot{\xi}(t)-\frac{\xi(t)}{T^2}=-\frac{4I \nu_0}{T^3}\delta(t).
\end{equation}
Here we have replaced $\nu(t)$ by
\begin{equation}\label{E_nu0}
\nu_0 = \nu(t=0)=\sqrt{1-\xi(t=0)};
\end{equation}
the corresponding criterion will appear shortly.
The solution of Eq.~(\ref{E_xi_delta}) is
\begin{equation}\label{E_xi_sol}
\xi(t) = \frac{2I \nu_0}{T} e^{-|t|/T}.
\end{equation}
In its turn,
\begin{align}\label{E_P}
P(t) =
\begin{cases}
\nu_0 e^{t/T}  & t < 0,\\
0                  & t > 0.
\end{cases}
\end{align}
What is left is to find $\nu_0$ by equating $\xi(t=0)$ from Eqs.~(\ref{E_nu0}) and (\ref{E_xi_sol}).
We obtain
\begin{equation}
\label{E_nu0_sol}
\nu_0 = \sqrt{\frac{I^2}{T^2} + 1} - \frac{I}{T}= \sqrt{V^2 + 1} - V.
\end{equation}
As one can check, $\displaystyle \max_t \xi(t)=\xi(0)<1$  with this $\nu_0$, as we assumed.
Figure \ref{fig:E_I} shows the analytic results for $q,p,\xi$ and $P$ versus time, along with $\delta^2(t)$.
The same figure also shows our numerical results for the same $I$ and $T$.
The corner singularity in $\xi(t)$ and the jump in $P(t)$, both observed at $t=0$, are the approximation price we have to pay for replacing $p(t)$ by the delta-function in the forcing term of Eq.~(\ref{E_xi}).
These singularities do not cause any problem in the action calculations which we now present.

With Eqs.~(\ref{E_xi_sol}) and (\ref{E_P}), the calculation of $S_\xi$ is straightforward:
\begin{equation} \label{E_S_xi}
S_\xi = \int_{-\infty}^{\infty} P \dot \xi \,dt = \frac{ \nu_0^2 I}{T} =V\left(\sqrt{V^2 + 1} - V\right)^2.
\end{equation}
The calculation of $S_q$, with $q$ and $p$ from Eq.~(\ref{E_qp}),  simplifies once we notice that the integral of $p\dot{q}$ over time is mostly gathered  in a narrow time interval of width $\sim 1/\nu_0$ around $t=0$.
Within this interval one can replace $\nu(t)$ by $\nu_0$ -- the same replacement as in Eq.~(\ref{E_xi_delta}) -- and obtain
\begin{equation} \label{E_S_q}
S_q = \int_{-\infty}^{\infty} p \dot q \,dt = \frac{2}{3} \nu_0^3 = \frac{2}{3} \left(\sqrt{V^2 + 1} - V\right)^3.
\end{equation}
The total action is
\begin{equation}\label{E_S}
\bar{S} = S_q + S_\xi = \frac{2}{3} \lt( 1 + V^2 \rt) ^\frac{3}{2} - V - \frac{2}{3} V^3.
\end{equation}
Equation (\ref{SSS}) with this $\bar{S}$ yields the MTE in this limit, up to a pre-exponential factor.

What is the validity domain of the adiabatic approximation which we have used?
An obvious condition is the strong inequality $\nu_0 T \ll 1$ which guarantees that the fast time $1/\nu_0$ is short compared with the slow time $T$.
Using Eq.~(\ref{E_nu0_sol}), one can reduce this strong inequality to
\begin{equation}\label{ineq1}
 T\gg \max\,(1,\sqrt{I}).
\end{equation}
This condition, however, is insufficient.
One also needs to demand that the variation of $\xi(t)$ during the fast time $1/\nu_0$ be small compared with each of the terms of Eq.~(\ref{C_HE_q}): for example, with $q^2$.
This criterion can be written as $\dot{\xi}(0)/\nu_0 \ll q^2$.
In view of Eqs.~(\ref{E_qp}) and~(\ref{E_xi_sol}) this criterion demands $I\ll (\nu_0T)^2$ which, after some algebra, boils down to $T\gg \max \, (I^{1/2}, I^{3/4})$.
[As one can check, the same criterion is required for the replacements of $\nu(t)$ by $\nu_0$ in Eqs.~(\ref{E_xi_delta}) and (\ref{E_S_q}).]
Combining this condition with Eq.~(\ref{ineq1}), we obtain the adiabaticity  criterion for the environmental noise:
\begin{equation}\label{E_cond}
 T \gg \max \,(1 , I^{3/4})
\end{equation}
or, in terms of the rescaled variance,
\begin{equation}\label{E_cond_var}
 T \gg \max \,(1 , V^3).
\end{equation}
For sufficiently large $T$ Eq.~(\ref{E_S}) agrees well with our numerical results, see the right solid line on Fig.~\ref{fig:DE_T_sim}.

\subsection{Weak Noise}
For sufficiently small $I$ the problem can be solved perturbatively.
Let us split the Hamiltonian (\ref{C_H}) into unperturbed and perturbed parts: $H=H_0+I H_1$, where
\begin{eqnarray}
  H_0 &=& 2 p^2 -  \lt( q^2-1 \rt) p  -   \xi p  -  \frac{1}{T} \xi P, \label{F_H0} \\
  H_1 &=& \frac{2}{T^2} P^2, \label{F_H1}
\end{eqnarray}
and $I$ serves as the small parameter.
Correspondingly,  $\bar{S}=\bar{S}_0+\Delta\bar{S}$, where the small correction $\Delta\bar{S}$  is proportional to $I$.

\subsubsection{Zeroth order}

The unperturbed, or zeroth-order problem is described by the Hamiltonian $H_0$, that is by Eqs.~(\ref{C_HE_q})-(\ref{C_HE_P}) with $I=0$.
The zeroth-order equation for $\dot{\xi}$ is $\dot{\xi}=-\xi/T$; its only acceptable solution is $\xi=\xi_0(t)= 0$: no environmental noise.
As a result, the zeroth-order equations (\ref{C_HE_q}) and (\ref{C_HE_p}) for $\dot{q}$ and $\dot{p}$ coincide with those without environmental noise, and their solutions, obeying the boundary conditions at $t=\pm \infty$, are
\begin{equation*}
q_0(t)=-\tanh t,\;\;\;\;\;\; p_0(t)=-\frac{1}{2\cosh^2 t}.
\end{equation*}
The action, contributed by the $(q,p)$ subsystem is $\bar{S}_0=2/3$.
Interestingly, the momentum $P_0(t)$, conjugate to $\xi_0(t)=0$, has a non-trivial behavior.
It is described by the equation
\begin{equation*}
\dot{P} - \frac{1}{T}P = - \frac{1}{2 \cosh^2 t}
\end{equation*}
whose solution, vanishing at $t=\pm \infty$, is
\begin{equation}\label{F_P0}
P_0(t) = \frac {e^{t / T}}{2} \int_t^\infty \frac{e^{- x / T}}{\cosh^2 x} dx.
\end{equation}
As $\xi_0(t)=0$, this ``ghost solution" does not contribute to the action, and $\bar{S}_0=2/3$, coming from the $(q_0,p_0)$ subsystem, yields Eq.~(\ref{MTE0}).
We will need the ``ghost solution", however, in the first order calculations which we now present.

\subsubsection{First order}
\begin{figure}[ht!]
\includegraphics[width=0.4\textwidth]{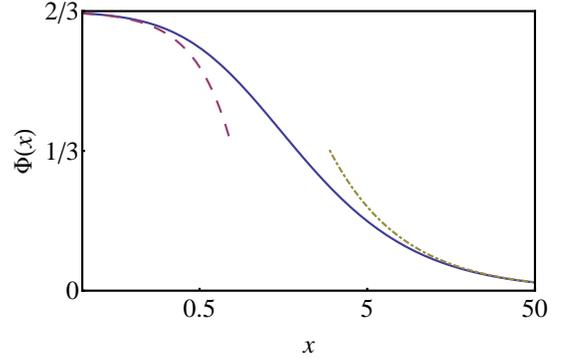}
\caption{(Color online)
The function $\Phi(x)$ from Eq.~(\ref{Phi}) and its small-$x$ (dashed) and large-$x$ (dot-dashed) asymptotics, $2/3-(8/15)x^2$ and $1/x$, respectively, shown on a semi-logarithmic scale. The  function  $\Phi(x)$ describes, for constant noise intensity, the dependence of the action on the correlation time in the limit of weak noise. If the noise \emph{variance} is held constant instead (not shown), the correlation-time dependence of the action becomes $T\Phi(T)$ which is an increasing function of $T$.}
\label{fig:F_Phi}
\end{figure}

The first-order correction to the action, $\Delta\bar{S}$, can be found by integrating $I H_1$ over the \emph{unperturbed} trajectories \cite{KMS,Schwartz,AKM}
\begin{equation*}
\Delta \bar{S}=-I\int_{-\infty}^\infty H_1[q_0(t),p_0(t),\xi_0(t),P_0(t)]\,dt.
\end{equation*}
Using Eqs.~(\ref{F_H1}) and~(\ref{F_P0}), we arrive at
\begin{eqnarray}
\!\!\!\!\!\!&&  \Delta \bar{S} = - \frac{2 I}{T^2} \int_{-\infty}^\infty P_0^2(t) dt \nonumber \\
\!\!\!\!\!\!&& = - \frac{I}{2T^2} \int_{-\infty}^\infty dt   \int_t^\infty dx \frac{e^{-x/T}}{\cosh^2 x}
\int_t^\infty dy \frac{e^{-y/T}}{\cosh^2 y}. \label{F_delta_S}
\end{eqnarray}
Evaluating this triple integral (see the Appendix for details), we obtain
\begin{equation}\label{F_dS}
\Delta \bar{S} = - I \,\Phi(T),
\end{equation}
where
\begin{equation} \label{Phi}
    \Phi(x)=\frac{1}{x^2}  \lt[ \frac{1}{2x} \varphi \lt( \frac{1}{2x} \rt) - x - 1 \rt],
\end{equation}
$\varphi (x)= d^2 \ln \Gamma(x)/dx^2$, and $\Gamma(x)$ is the gamma-function.
$\varphi$ is the so called trigamma function: a special case of  the polygamma function \cite{Abramowitz}.
The function $\Phi(x)$ is plotted on Fig.~\ref{fig:F_Phi} along with its  small- and large-$x$ asymptotics.

\begin{figure}[ht!]
\includegraphics[width=0.45\textwidth]{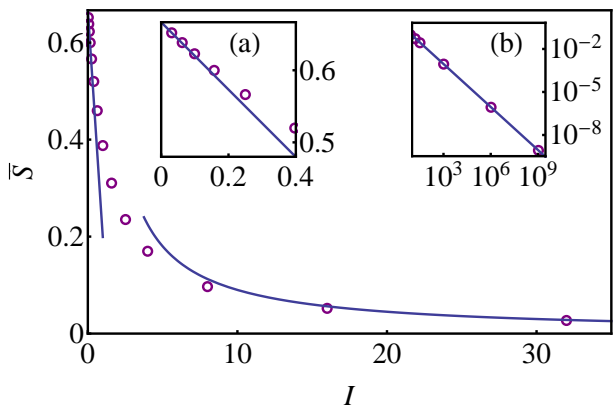}
\caption{(Color online) Analytic (lines) and numerical (circles) results for the action $\bar{S}$ versus $I$ for $T=1$. As the noise intensity increases, the mean time to extinction decreases as expected.
The small-$I$ asymptotic is the prediction of weak-noise theory, Eq.~(\ref{S1}) [it is also shown in inset (a)].
For $I \gg \max(1,T)$  $\bar{S}$ behaves as $\bar{S}=g(T)/I$, see subsection \ref{strong}.
The function $g(T)$ is only known analytically for $T \ll 1$ and for $1\ll I^{3/4}\ll T \ll I$. The right line is $\bar{S}=0.9/I$,
that is $g(T=1)\simeq 0.9$. Inset (b): $\bar{S}$ versus $I$ for very large $I$ is displayed on a log-log scale.
The slope of this asymptotic is $-1$ as expected.}
\label{fig:F_I_sim}
\end{figure}

Altogether, our weak-noise result for the  action is
\begin{equation}\label{S1}
\bar{S}=\frac{2}{3} -  \frac{I}{T^2} \lt[ \frac{1}{2T} \varphi \lt( \frac{1}{2T} \rt) - T - 1 \rt].
\end{equation}
Using the small- and large-$x$ asymptotics of $\varphi(x)$, we can obtain simple formulas for  $\bar{S}$ for short- and long-correlated noise
\begin{align}\label{F_dS_lim}
\bar{S} \simeq
\begin{cases}
 \dfrac{2}{3}\left(1-I  + \dfrac{4 IT^2}{5} \right),  & T \ll 1,\\
\dfrac{2}{3}-\dfrac{I}{T},                                               & T \gg 1.
\end{cases}
\end{align}
As one can easily check, the $T\ll 1$ asymptotic in Eq.~(\ref{F_dS_lim}) coincides with the $I\ll 1$ asymptotic of Eq.~(\ref{D_S}) obtained for the short-correlated noise.
In its turn, the $T\gg 1$ asymptotic in Eq.~(\ref{F_dS_lim}) coincides with the $V\ll 1$ asymptotic of Eq.~(\ref{E_S}) obtained for the long-correlated noise.
Figure \ref{fig:F_I_sim} shows a comparison of Eq.~(\ref{S1}) with our numerical results for $T=1$.
One can see good agreement for sufficiently small $I$.

Now we can determine the validity domain of the weak-noise approximation by demanding the strong inequality $\Delta \bar{S}\ll \bar{S}_0$, or simply $\Delta \bar{S}\ll 1$.
For $T\lesssim 1$ we have $\Delta S \sim I$, whereas for $T \gtrsim  1$ we obtain   $\Delta S \sim I/T$.
Therefore, the weak-noise approximation holds when $I \ll \max\,(T,1)$ or, in terms of the rescaled variance, $V \ll \max\,(1/T,1)$.

\subsection{Strong noise}
\label{strong}
For strong environmental noise we only have partial results which,  as we will see shortly, are not new. How does $\bar{S}$ depend on $I$ for very large $I$? Here the demographic noise becomes negligible compared with the environmental noise. Therefore, the parameter $K$ must drop from the exponent $KS$ of the MTE in Eq.~(\ref{MTE}). This can only happen if $S$, and therefore $\bar{S}=S/\delta_0^3$ behaves as $\bar{S}=g(T)/I$.
We can extract $g(T)$ from our analytic results in two different domains. For the short-correlated noise, $T\ll 1$, we can expand Eq.~(\ref{D_S}) at $I\gg 1$. For the long-correlated noise, we can use the large-$V$ asymptotic of Eq.~(\ref{E_S}). These procedures yield
\begin{align}\label{ftwodom}
g(T) \simeq
\begin{cases}
 \frac{2}{3}+\frac{8T^2}{15},  & I\gg 1,\;\;T \ll 1,\\
\frac{T}{4},                                               & 1\ll I^{3/4}\ll T \ll I.
\end{cases}
\end{align}
These asymptotics can be compared with those obtained in 1989 by Bray and McKane  \cite{BM}. They investigated escape of an overdamped particle from a smooth potential well $U(x)$ solely due to an (Ornstein-Uhlenbeck)  extrinsic noise with correlation time $\tau_c$ and intensity $D$. The absence of intrinsic noise in their setting  corresponds to the limit of strong environmental noise in ours. Bray and McKane \cite{BM} presented  their result for the mean time to escape as $\ln \tau \simeq s/D$. To go over to our notation, we use Eq.~(\ref{I_def}) to express $D=2I/K$. As a result, $\ln \tau \simeq K s/(2I)$, and our $g(T)$ is related to their $s$ as $g(T)=s/(2\delta_0^3)$.

For short-correlated noise Bray and McKane arrived at
\begin{eqnarray}
  s &=& U(a)-U(b) + \tau_c^2 \int_a^b dx \, [U^{\prime\prime}(x)]^2 \,U^{\prime}(x) \nonumber \\
  &-& \tau_c^4 \int_a^b dx \, [U^{\prime\prime\prime}(x)]^2 \,[U^{\prime}(x)]^3 + \mathcal{O}(\tau_c^6),
  \label{BMshort}
\end{eqnarray}
where the fixed points $a$ and $b$ correspond to our $1+\delta$ and $1-\delta$, respectively. Putting $U(q)=U_0(q)=(q-1)^3/3-\delta_0^2 q$, limiting ourselves only to the leading correction ${\cal O}(\tau_c^2)$, and evaluating the integral in Eq.~(\ref{BMshort}), we obtain
$$
s=\left(\frac{4}{3}+\frac{16T^2}{15}\right) \delta_0^3,
$$
where $T=\delta_0 \tau_c$.  This yields the first line in our Eq.~(\ref{ftwodom}).

For long-correlated noise Bray and McKane \cite{BM} obtained
\begin{equation}\label{BMlong}
s=\frac{\tau_c}{2}\,\left[U^{\prime}(d)\right]^2,
\end{equation}
where $d$ is the inflection point of the potential $U(q)$, located between the points $a$ and $b$. For our $U_0(q)$ one has $d=1$, and Eq.~(\ref{BMlong}) yields $s=\delta_0^3 T/2$ which leads to the second line in our Eq.~(\ref{ftwodom}).

For $T\sim 1$ analytic progress is difficult, as was already noticed in Ref.~\cite{BM}. Still, $g(T)$ can be found numerically, see also Ref.~\cite{BM}.  For example, we found that $g(1)\simeq 0.9$, see Fig.~\ref{fig:F_I_sim}. Finally, the strong-noise limit corresponds to $I\gg \max(1,T)$, or $V\gg \max(1/T,1)$.

\subsection{Phase diagram}
\begin{figure}[ht!]
\includegraphics[width=0.3\textwidth]{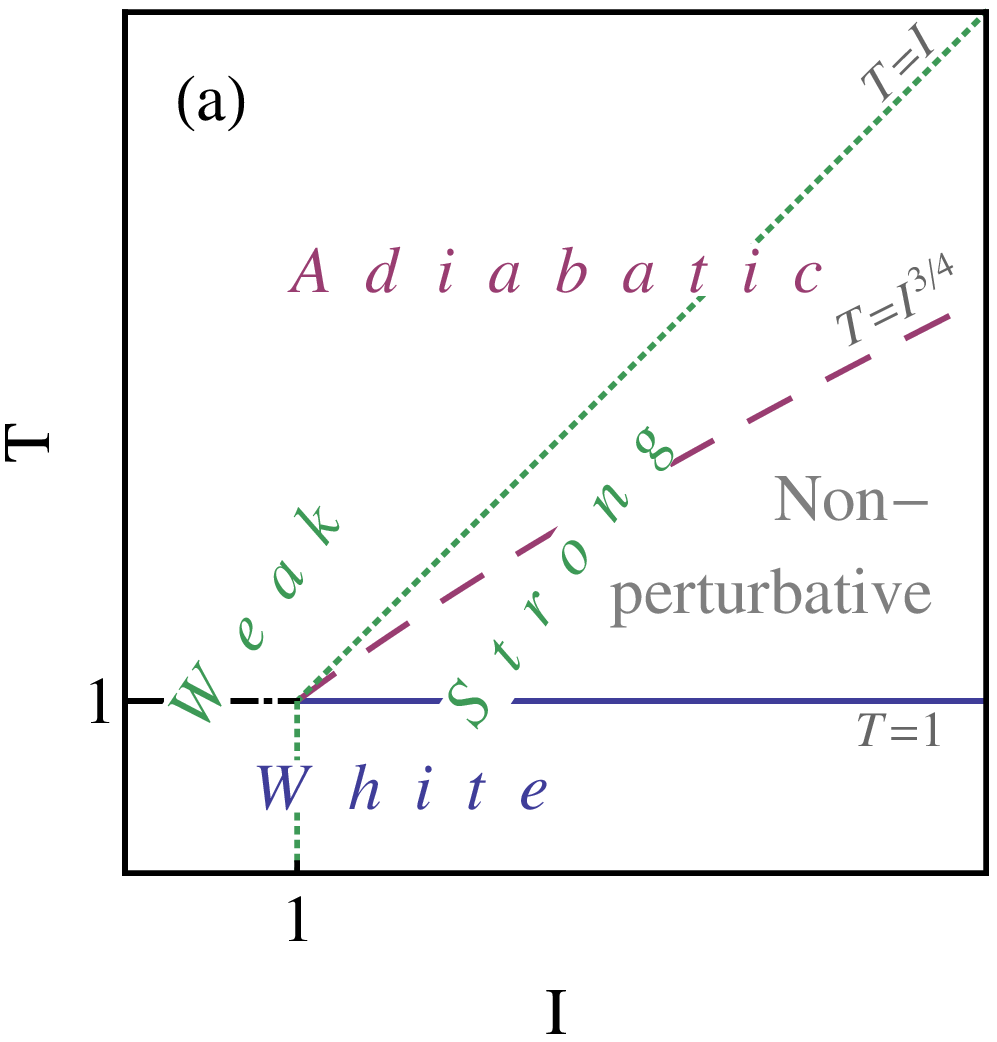}
\includegraphics[width=0.3\textwidth]{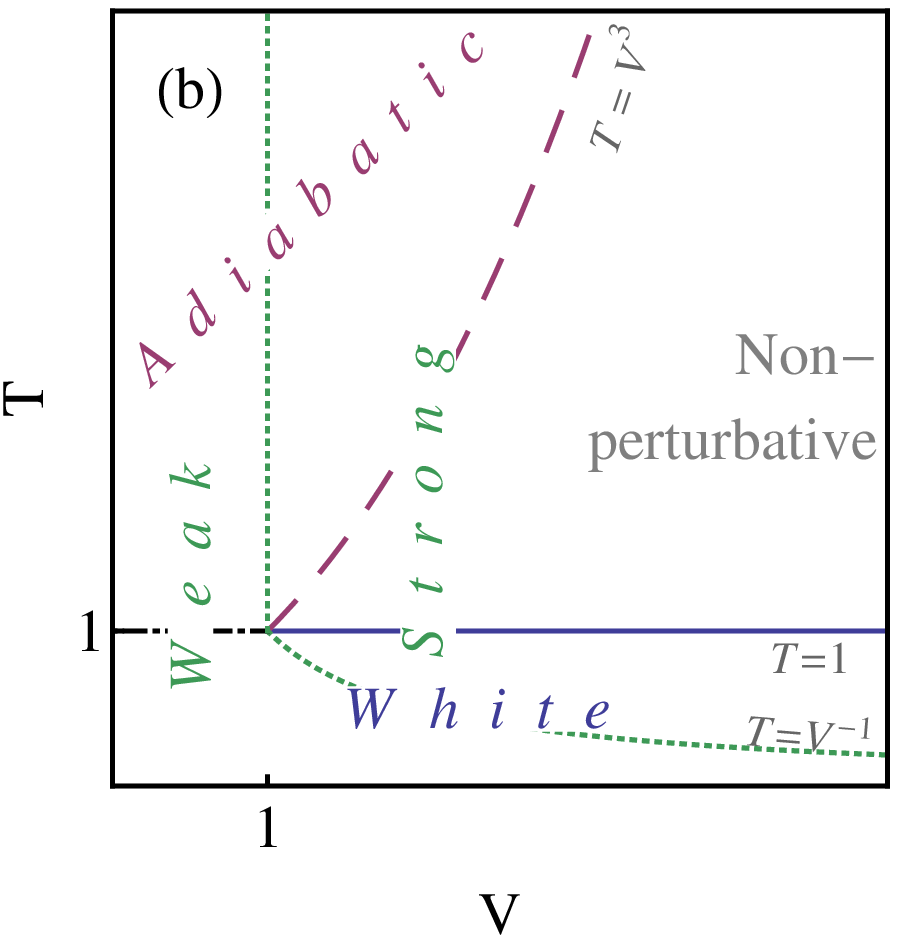}
\caption{
(Color online) Phase diagram of the system
on the $(I,T)$ (a)  and
$(V,T)$ (b) planes. It shows the
 validity regions of our results for $\bar{S}\sim \ln \tau/(K\delta_0^3)$ summarized in Table \ref{table:S}.}
\label{fig:PD}
\end{figure}

Table \ref{table:S} summarizes our main analytic results for $\bar{S} \simeq (K\delta_0^3)^{-1} \ln \tau$ in different regions of the parameter plane $(I,T)$. The regions themselves make a ``phase diagram" which is shown in Fig.~\ref{fig:PD} on the $(I,T)$  and $(V,T)$ planes.

\begin{table}[thb]
\centering
\begin{tabular}{c|c|c}
Noise       &   Equation                &  $\bar{S}$
\\
\hline
\hline

Almost white       &  (\ref{D_S})              & $\frac{2}{3(I+1)}
                                            \left(1+\frac{4}{5}\frac{I T^2}{I+1}\right) $
\\
\hline

Adiabatic   &  (\ref{E_S})              & $\frac{2}{3} [1 + \left(\frac{I}{T}\right)^2]^{3/2}
                                                             - \frac{I}{T} -   \frac{2}{3} \left(\frac{I}{T}\right)^3 $
\\
\hline

Weak        &  (\ref{S1})               & $ \frac{2}{3} - \frac{I}{T^2} \left[ \tfrac{1}{2T} \varphi(\tfrac{1}{2T}) - T - 1 \right] $

\\
\hline
\end{tabular}
\caption{Action $S$ in different parameter regions}
\label{table:S}
\end{table}

The validity domains of our results for the purpose of evaluation of the MTE can be found in each particular case by demanding that $K\delta_0^3 \bar{S}\gg 1$. In the cases where we calculated a small sub-leading term $\bar{S}_1$, a more stringent condition $K\delta_0^3 \bar{S}_1\gg 1$ is required. These criteria always hold for sufficiently large $K$ and small $D$.

\section{Summary}

We have evaluated the mean time to extinction (MTE) of a long-lived and well-mixed  isolated population caused by an interplay of colored environmental noise and (effectively white) demographic noise. We assumed that the population exhibits a strong Allee effect. We have obtained analytic results in the limits of short-correlated, long-correlated and (relatively) weak environmental noise, see Table 1. We have also established the validity domains of white, adiabatic, weak and strong noises on the parameter plane. As in the absence of the Allee effect,  even a relatively weak environmental noise leads to an exponentially large reduction in the MTE. For a relatively strong environmental noise, this effect becomes dramatic. We have found, in the different limits, the most likely path of the population to extinction and the optimal environmental fluctuation (OEF) that mostly contributes to this path.

For a relatively strong and short-correlated environmental noise the OEF temporarily changes the sign of the difference between the birth and death rates of the population. For long-correlated noise the OEF is such that this difference remains positive at all times, for any noise intensity.

This theory is immediately extendable, close to the saddle-node bifurcation, to population transitions,  due to a combined action of demographic and environmental noise, in two additional settings. The first setting is population explosion. The second is population switches between two non-empty states each of which, at the deterministic level, is linearly stable. Without environmental noise, these problems were considered in Refs. \cite{DykmanRoss,EK,Doering,explosion,EscuderoK}.

Finally, it would be overly optimistic to hope that an analysis of a simple model which we presented here will resolve the long-time debate in population biology on ``whether and under which conditions red noise increases or decreases extinction risk compared with uncorrelated (white) noise" \cite{Schwager}. Still, we believe that this analysis is a step toward resolving this debate.

\section*{ACKNOWLEDGMENTS}

EYL is grateful to Vadim Asnin for helpful discussions.
This work was supported by the Israel Science Foundation (Grant No. 408/08) and by the US-Israel Binational Science Foundation (Grant No. 2008075).

\section*{APPENDIX: CALCULATION OF THE WEAK-NOISE INTEGRAL}
\renewcommand{\theequation}{A\arabic{equation}}
\setcounter{equation}{0}

Here we present some details of the calculation of the triple integral in Eq.~(\ref{F_delta_S}).
Let us change the integration order by moving the integral over $t$ to the innermost position.
Since the integration is over $\{(t,x,y): x>t, y>t\}$,
the $x$- and $y$-integration domains are $(-\infty,\infty)$,
whereas for any $x,y$ the $t$-integration
is from $-\infty$ to $u = \min(x,y)$. The $t$-integration yields $(T/2) \,e^{2 u/T}$.
Now we split the integration domain of $y$ into two sub-domains:
$y<x$ (where $u=y$) and to $y>x$ (where $u=x$). We obtain
\begin{align*}
\Delta S &= -\frac{I}{4T} \int_{-\infty}^\infty dx\, \frac{e^{-x/T}}{\cosh^2 x}
\begin{aligned}[t]
\bigg[ & \int_{-\infty}^x      dy \,\frac{e^{y/T}}{\cosh^2 y} \\
     + & \int_x        ^\infty dy \,\frac{e^{(2x-y)/T}}{\cosh^2 y}\bigg]
\end{aligned}
\\
 &= -\frac{I}{2T}
	\int_{-\infty}^\infty dx \int_{-\infty}^x      dy\, \frac{e^{(y-x)/T}}{\cosh^2 x \cosh^2 y}\,.
\end{align*}

\noindent
Next we shift $y$: $\bar{y}=y-x$. Omitting the overbars
and changing the integration order, we obtain
\begin{align*}
\Delta S
&= -\frac{I}{2T}
    \int_{-\infty}^0      dy \,e^{y/T}
    \int_{-\infty}^\infty \frac{dx}{\cosh^2 x \cosh^2 (y+x)}  \\
&= - \frac{2I}{T}
    \int_{-\infty}^0      dy \,e^{y/T}
    \lt( \frac{y \cosh y}{\sinh^3 y} - \frac{1}{\sinh^2 y} \rt) \\
&= -\frac{I}{2 T^3} \lt[ -2 T (T+1) + \varphi\left(\frac{1}{2 T}-1\right) - \frac{4 T^2}{(1 - 2 T)^2} \rt] \\
&= -\frac{I}{T^2} \left[    \frac{1}{2 T} \varphi\left(\frac{1}{2 T}\right) - T - 1 \right],
\end{align*}
where $\varphi(z) = d^2 \ln\Gamma(z) / d z^2$ is the trigamma function, and  we have used the identity
$\varphi(z-1)-(z-1)^{-2}=\varphi(z)$ \cite{Abramowitz}.


\begin{thebibliography}{99}
\bibitem{shift} These time dependences include an arbitrary time shift. It is inconsequential in
this problem, so we put it to zero for brevity.
\end{thebibliography}
\end{document}